\documentclass[epsfig,12pt,onecolumn]{article}
\usepackage{amsfonts}
\usepackage{amssymb}
\usepackage{amsmath}
\usepackage{multicol}
\usepackage{graphicx}
\usepackage{float}
\usepackage{caption}

\input{tcilatex}
\makeatletter \@addtoreset{equation}{section}

\def \be{\begin{equation}}
\def \ee{\end{equation}}
\def \bea{\begin{eqnarray}}
\def \eea{\end{eqnarray}}

\newcommand{\nc}{\newcommand}
\nc{\al}{\alpha} \nc{\bib}{\bibitem} \nc{\la}{\lambda}
\nc{\C}{\mbox{\hspace{1.24mm}\rule{0.2mm}{2.5mm}\hspace{-2.7mm} C}}
\nc{\R}{\mbox{\hspace{.04mm}\rule{0.2mm}{2.8mm}\hspace{-1.5mm} R}}

\begin{document}

\title{\textbf{Implications of the NANOGrav results for primordial black
holes and Hubble tension}}
\author{M. Bousder$^{1}$\thanks{%
mostafa.bousder@fsr.um5.ac.ma}, A.Riadsolh$^{3}$, A. El Fatimy$^{4,5}$, \\
M. El Belkacemi$^{3}$ and H. Ez-Zahraouy$^{1,2}$ \\
%EndAName
$^{1}${\small Laboratory of Condensend Matter and Interdisplinary Sciences,
Department of physics,}\ \\
{\small Faculty of Sciences, Mohammed V University in Rabat, Morocco}\\
$^{2}${\small CNRST Labeled Research Unit (URL-CNRST), Morocco}\\
$^{3}${\small Laboratory of Conception and Systems (Electronics, Signals and
Informatics)}\\
\ {\small Faculty of Sciences, Mohammed V University in Rabat, Morocco}\\
$^{4}${\small Central European Institute od Technology,}\\
\ {\small CEITEC BUT, Purky\v{n}ova 656/123, 61200 Brno, Czech Republic}\\
$^{5}${\small Departement of Physics, Universit\'{e} Mohammed VI
Polytechnique, Ben Guerir 43150, Morocco}\\
}
\maketitle

\begin{abstract}
The purpose of this work is to investigate the formation and evaporation of
the primordial black holes in the inflationary scenarios. Thermodynamic
parameters such as mass, temperature and entropy are expressed in terms of
NANOGrav frequency. By numerical calculations we show that the constraint on
the mass range $10^{-5}kg-10^{50}kg$ is well confirmed. We discuss the
relation between the redshift and the probability for gravitational wave
source populations. A new parameter associated with the frequency and Hubble
rate is presented, by which for the spectral index $n_{s}\approx 0.996$ and
the Hubble constant $H_{0}\approx 67.27km.s^{-1}.Mpc^{-1}$, the effective
Hubble constant is calculated to be $H_{eff,0}\approx
73.24km.s^{-1}.Mpc^{-1} $ which is compatible with the observational data.
We make a comparison between the Hubble tension and the primordial
perturbations and the expression of the mass loss rate, chemical potential
and central charge needed to describe the Hawking evaporation will be
established.
\end{abstract}

\section{Introduction}

There has been a recent surge of interest in the 15 yr of pulsar timing data
of stochastic gravitational wave (GW) signal \cite{NANO0,NANO00}, published
by the North American Nanohertz Observatory for Gravitational Waves
(NANOGrav) collaboration, In \cite{NANO1}, the authors have been working on
12.5 yrs of pulsar timing data. Recently, several gravitational wave models
have been proposed \cite{NANO2,NANO3}, in an attempt to explain the NANOGrav
results. The observed signal could be attributed to a stochastic
gravitational wave background originating from primordial black holes (PBHs)
formed during inflation \cite{NANO4}. Conversely, alternative studies
propose cosmic strings as the source of these wave phenomena. \cite{NANO5}.
The potential GW signal can be generated by the power-law of abundance $%
\Omega _{GW}\propto f^{\zeta }$. The NANOGrav observes the exponent $\zeta
\in \left( -1.5,0.5\right) $ at a frequency $f=5.5nHz$. The LIGO
collaboration and Virgo collaboration \cite{PBH0,PBH01} detect the GWs from
the black holes. The PBHs can be taken as dark matter candidates and can
also have formed in the early Universe \cite{PB0}, in this regard, we
assumed that the dark matter halo is composed of these PBHs in the edges of
the galaxies. From the constraints on PBHs, the mass range of PBH is $%
10^{-5}g-10^{50}g$ \cite{PBH02}. The PBH mass that can form dark matter is
around $10^{20}g$ \cite{PBH03}. The PBHs formed before the end of the
radiation-dominated era and are not subject to the big bang nucleosynthesis
(BBN), such that more than 5$\%$ of the critical density forms the baryons
\cite{PBH04}. Inflation predicted a stochastic background of GWs over a
broad range of frequencies, which have a direct relationship with cosmic
microwave background (CMB) measurements \cite{R21}.\ Also, they find a
constraint on the spectral index as $n_{s}>0.9$, which is in good agreement
with the the Planck 2018 results ($n_{s}^{0}\approx 0.965\pm 0.004$) \cite%
{L2}.\newline
On the other hand, the Hubble tension is one of the most intriguing problems
in cosmology today. Local measurements of the current expansion rate of the
Universe based on Type Ia Supernovae (SNeIa) observations from the Hubble
Space Telescope (HST), give values that are grouped around $H_{0}\simeq
\left( 73.24\pm 1.74\right) km.s^{-1}.Mpc^{-1}$ \cite{L1}. While, the value
of the Hubble constant using the CMB measurements comes out at around $%
H_{0}\simeq \left( 67.27\pm 0.66\right) km.s^{-1}.Mpc^{-1}$ \cite{L2}, and
the deviation from local measurements is now 4$\sigma $ or more. Among of
the more successful solutions to the Hubble tension is the addition of an
early dark energy component \cite{PL1}\ or primordial magnetic fields \cite%
{PL2}. There are also several works that have evaluated this problem by
constraints. There has been lots of discussions about new physics behind
this tension in the form of modifications or additions of energetic
components to the standard cosmological model \cite{PLN,PLM}. Recently, M.
R. Visser \cite{R3} proposed a novel holographic central charge $C\sim
L^{d-2}$, which has attracted great attention. He finds that the Euler
equation is dual to the generalized Smarr formula for black holes in the
presence of a negative cosmological constant. This theory sheds light on the
holographic and thermodynamic aspects of black holes. In comparison, authors
in \cite{GG1,GG2} have been proposing a restricted phase space approach, in
which the pressure $P$\ and the volume $V$\ are absent in the thermodynamic
description for black holes. While in \cite{GG3} through the Holographic
Principle, the limit values of the mass density of PBHs formed in the early
universe can be obtained. Motivated by these and specifically by the work
\cite{NANO4}, the purpose of the present letter is to introduce an
interpretation of the Hubble tension from NANOGrav signal in comparison with
the PBHs formation and rotation. We present a new paradigm for explaining
the stochastic gravitational wave background from the PBHs. We also aim to
connect the horizon mass with comoving frequency of the perturbation. The
goal of this paper is to show that there is more information about the early
universe hidden in the thermodynamic properties of PBHs.

This paper is organized as follows. In Sec. 2, we analyze the stochastic
gravitational wave background from the PBHs and will establish relations for
thermodynamical parameters of PBHs. In Sec. 3, we calculate the slow roll
parameter and spectral index in inflationary scenarios for PBH formation.
Also, we explore the Hubble tension from the spectral index. In Sec. 4, the
formation of PBHs and their subsequent evaporation and their relationship
with a new parameter associated with the Hubble rate are studied. In Sec. 5,
the gauge/gravity duality aspect of the PBHs by the chemical potential and
central charge are discussed and we conclude our findings in Sec. 6.

\section{Method}

In order to study the stochastic gravitational wave background from the PBHs
generated during inflation, it is necessary to understand the mechanism by
which there is an accelerated expansion of the Universe. We work in the
framework with units $c\neq G\neq \hbar \neq 1$. The metric of a spatially
flat homogeneous and isotropic universe in the Robertson Walker spacetime
with perturbations is given by \cite{L03}:
\begin{equation}
ds^{2}=\left( ic^{2}dt\right) ^{2}+a^{2}(t)\left( \delta _{ij}+h_{ij}\right)
dx^{i}dx^{j},  \label{a1}
\end{equation}%
where $h_{ij}$ is a $3\times 3$ symmetric matrix of the perturbations of
spacetime, $(i,j=1,2,3)$ and $a\left( t\right) $ is the scale factor that
describes the relative expansion of the Universe. We define the Hubble
parameter as $H=\dot{a}/a$ with $\dot{a}$ denoting derivative of $a$ with
respect to time. In the expanding Robertson Walker space-time the scale
factor $a\left( t\right) $ is related to the wave length $\lambda $ and the
comoving wave number $k$ by \cite{La03,L3}:%
\begin{equation}
k\lambda =2\pi a.  \label{a2}
\end{equation}%
To make progress, let us introduce the frequency $f=kc/2\pi $, which is
associated with the stochastic gravitational wave background. From Eq. (\ref%
{a2}) we have%
\begin{equation}
a(f)=2\pi f\left( \frac{\lambda }{2\pi c}\right) .  \label{a3}
\end{equation}%
Using this and from the complex conformal transformation \cite{La03,La3}, we
assume that the scale factor associated with gravitational wave background
has the following form
\begin{equation}
a\left( f\right) \equiv a_{0}e^{2\pi f\tau }.  \label{a5}
\end{equation}%
where $a_{0}$\ is a constant and $\tau $ is the Euclidean time (Wick
rotation $it\rightarrow \tau $) \cite{L4,L5}. One should note that to
determine the temperature associated with respect to e-folds, one must first
determine the entropy of the system in terms of GWs frequency. The entropy $%
s $ is expressed as a function of the number of e-folds $N=\ln \left(
a\left( t\right) /a_{0}\right) $ remaining to the end of the inflation
period \cite{R0}. The number of e-folds measures the amount of inflation
that occurs. Note that the entropy of the PBH can be modeled by a
logarithmic function \cite{R00,R0},%
\begin{equation}
s=k_{B}\ln \frac{a\left( f\right) }{a_{0}}=k_{B}N,  \label{p3}
\end{equation}%
where $N$ is the number of e-folds between the horizon exit and the end of
inflation. The quantity $a\left( f\right) /a_{0}$ at the moment of the PBH
formation must be greater than $1$. According to the positivity of $s$, we
find that $a\left( f\right) >a_{0}$. At the end of inflation ($N\approx 0$),
the entropy vanishes. The Wick rotation $it\rightarrow \tau $ \cite{L4}
links statistical physics and quantum mechanics by replacing the inverse
temperature $1/k_{B}T$ with a periodicity of imaginary time $\tau /\hbar $.
So, from Eq. (\ref{a5},\ref{p3}) the entropy is given by%
\begin{equation}
s\left( f\right) =\frac{2\pi \hbar f}{T},  \label{a6}
\end{equation}%
This entropy is expressed as a function of $f$. Using the Boltzmann type of
entropy $s=-k_{B}\ln \mathcal{P}\left( t\right) =-k_{B}\ln \frac{a_{0}}{%
a\left( f\right) }$, where%
\begin{equation}
\mathcal{P}\left( f\right) =e^{-2\pi f\tau }\leq 1,  \label{p}
\end{equation}%
is the probability of the redshift $z$ and is a solution of an evolution
Fokker-Planck equation, leading to its current distribution. The probability
distribution is taken to have the non-Gaussian form. This probability is in
good agreement with the estimation of the sensitive volume \cite{L10}, for
GW source populations at redshift. For the power law form of the scale
factor $a\left( t\right) =a_{0}t^{q}$, the probability distribution reduces
to $\mathcal{P}\left( t\right) =t^{-q}\leq 1$. In the dark-energy-dominated
era ($a\left( t\right) =a_{0}e^{H_{0}t}$), the evolution of the probability
is $\mathcal{P}\left( t\right) =e^{-H_{0}t}\leq 1$ with $2\pi f\tau =H_{0}t$%
. Considering $t\rightarrow \infty $, the probability distribution becomes
zero, i.e. the boundary condition $\mathcal{P}\left( t\right) =0$. The
temperature $T$ determines how fast probability decays with energy.\newline
In addition, from Eq. (\ref{p3}) and Eq. (\ref{a6}) it is evident that
\begin{equation}
T=\frac{2\pi \hbar f}{k_{B}N},  \label{p4}
\end{equation}%
where $T$ is the temperature of the system, which depends on the frequency $%
f $ of the stochastic gravitational wave background. At the end of inflation
($N\approx 0$), the temperature tends to infinity. Eq. (\ref{p4})
corresponds with Wien's displacement law. The temperature fluctuations are
projected as anisotropies on the cosmic microwave background (CMB) sky \cite%
{M1} by the number of e-folds $N$. The temperature $T$ grows if the
conformal time grows quickly. The radiation energy density today is given by
$\Omega _{r}h^{2}=4.18\times 10^{-5}$ for $T_{CMB}=2.725K$ \cite{L7}. From\
Eq. (\ref{p3}) and Eq. (\ref{p4}) we note that $T=\frac{2\pi \hbar f}{s}%
\propto \frac{1}{M}$ with $M$ being PBH mass, which is in good agreement
with the temperature of black holes radiation \cite{R5}. The black holes
radiate thermally with temperature $T=\frac{\hbar c^{3}}{8\pi GMk_{B}}$ \cite%
{R5}, which leads to the PBH mass:%
\begin{equation}
M\left( f\right) =\frac{c^{3}}{G}\frac{N}{16\pi ^{2}f}.  \label{p5}
\end{equation}%
We notice that $M\propto 1/f$, which is compatible with the result found in
\cite{NANO4}. Since the ratio $N/f$ has the dimension of time, which
corresponds to the order of PBHs mass $M\sim \frac{c^{3}}{G}t$ \cite{PBH02}.
On the other hand, the PBHs with the mass less than $10^{15}g$ would have
completely been evaporated. Non-evaporating PBHs are characterized by $%
M>10^{15}g$. Also, the mass range of PBH is $10^{-5}g-10^{50}g$ \cite{PBH02}%
. The e-folds before the end of inflation for the formation of the visible
part of the Universe is $N=60$ \cite{L12,L13,L14}. The PBHs that have been
produced in the early universe are referred to as the production of the
galaxies \cite{LL}. By using Eq. (\ref{p5}), we show that $f$ is the
frequency of the waves emitted by the PBHs. One interesting point of
discussion could then be the following. First, we study the cases between
the moments before the end of inflation ($N=60$) and at the end of inflation
($N=0$). As is very clear from the table \ref{t1}, we notice that the
relation (\ref{p5}) well confirms the constraints on the mass range of PBH
\cite{PBH02} in addition to obtaining the value $f=5.5nHz$ for the frequency
which is detected by the NANOGrav collaboration.
\begin{table}[H]
\begin{equation*}
\begin{tabular}{ccc}
\hline
$N$ & $M\left( g\right) $ & $f\left( nHz\right) $ \\ \hline\hline
$10^{-47}$ & $10^{-2}$ & $5.500$ \\
$2.15\times 10^{-30}$ & $10^{15}$ & $5.500$ \\
$10$ & $4.65\times 10^{45}$ & $5.507$ \\
$20$ & $9.30\times 10^{45}$ & $5.507$ \\
$30$ & $1.40\times 10^{46}$ & $5.487$ \\
$40$ & $1.86\times 10^{46}$ & $5.507$ \\
$50$ & $2.3\times 10^{46}$ & $5.567$ \\
$60$ & $2.75\times 10^{46}$ & $5.587$ \\
$60$ & $2.78\times 10^{46}$ & $5.527$ \\
$60$ & $2.78\times 10^{47}$ & $1.536$ \\ \hline
\end{tabular}%
\end{equation*}%
\caption{Numerical estimate of the values of the PBH mass, according to the
frequency $f$ and the interval of $N$ between $10^{-47}$ and $60$. Such as $%
f=2.561\times 10^{33}\frac{N}{M}.$}
\label{t1}
\end{table}
Recently, the NANOGrav collaboration published an analysis of 12.5 yrs of
pulsar timing data of stochastic gravitational wave signal \cite{NANO1}.
This signal may be interpreted as a stochastic gravitational wave background
from the primordial black holes (PBHs) generated during inflation \cite%
{NANO4}. The potential GW signal can be generated by the power-law of
abundance $\Omega _{GW}\propto f^{\zeta }$. The NANOGrav observes the
exponent $\zeta \in \left( -1.5,0.5\right) $ at a frequency $f=5.5nHz$. We
assume that $f$ is the frequency of the stochastic gravitational waves which
are emitted in the early Universe. During phase transitions in the early
universe at frequency $f=5.5nHz$, from Eq. (\ref{p4}) we obtain
\begin{equation}
N=\frac{T_{0}}{T},
\end{equation}%
where $T_{0}=2.64\times 10^{-19}K$. Temperature $T$ must be in the order of $%
T_{0}$, which shows that the PBHs are very cold, which is in good agreement
with the cold dark matter (CDM) model. The particles move slowly compared to
the speed of light in the CDM. The scale factor describes the relative
expansion of the Universe $a\left( t\right) $. The scale factor $a\left(
t\right) $ changes according to the chronology of the Universe. Next, we
assume that $H=q/t$, then, the scale factor has the specific form $a\left(
t\right) =a_{0}t^{q}$. In this case, we recall that in the
radiation-dominated era, we have $a\left( t\right) \propto t^{1/2}$. On the
other hand, during the matter-dominated era, the scale factor behaves as $%
a\left( t\right) \propto t^{2/3}$. Note that during the
dark-energy-dominated era, the evolution of the scale factor is $a\left(
t\right) =a_{0}e^{H_{0}t}$.
\begin{table}[H]
\begin{center}
\begin{equation*}
\begin{tabular}{ccc}
\hline
Era & $a\left( t\right) \propto $ & $\Omega _{GW}\left( f\right) \propto $
\\ \hline\hline
Radiation-dominated era & $t^{1/2}$ & $f^{1/2}$ \\
Matter-dominated era & $t^{2/3}$ & $f^{-3/2}$ \\ \hline
\end{tabular}%
\end{equation*}%
\end{center}
\caption{Two types of the stochastic background of gravitational waves
according to the chronology of the Universe.}
\label{t2}
\end{table}
Let us briefly discuss the different values of $\Omega _{GW}\left( f\right) $
according to the chronology of the Universe in Table \ref{t2}. If the PBHs
are more dynamic, then the $\Omega _{GW}$ parameter will be more important
in the early Universe,\ which means that the case $\Omega _{GW}\propto
f^{1/2}$ is the one that describes the radiation-dominated era. While the
case $\Omega _{GW}\propto f^{-3/2}$ describes the matter-dominated era.

\section{PBHs formation and Hubble tension}

The inflationary scenarios for PBHs formation were proposed in \cite{R16,R17}%
. Now, when we combine Eq. (\ref{p5}) with the PBH mass $M\sim \frac{c^{3}}{G%
}t$ together, we find\
\begin{equation}
f=\frac{N}{16\pi ^{2}t}.  \label{in1}
\end{equation}%
Using Eq. (\ref{in1}) and $\dot{N}=-H$ with $H=q/t$, we find the slow roll
parameter $\epsilon _{H}=-\dot{H}/H^{2}\ll 1$ for constant $N$:%
\begin{equation}
\epsilon _{H}=-\frac{1}{16\pi ^{2}qf^{2}}\left( N\dot{f}+Hf\right) ,
\label{in2}
\end{equation}%
yielding
\begin{equation}
\dot{\epsilon}_{H}=\frac{1}{16\pi ^{2}qf^{2}}\left[ 2\left( N\frac{\dot{f}%
^{2}}{f}+H\dot{f}\right) -\left( N\ddot{f}-\ddot{N}f\right) \right] .
\label{in3}
\end{equation}%
From this model, the slow-roll\ parameter $\eta =\epsilon _{H}-\dot{\epsilon}%
_{H}/2H\epsilon _{H}$ \cite{L15} is given by%
\begin{equation}
\eta =\frac{1}{2H\left( N\dot{f}+Hf\right) }\left[ 2\left( N\frac{\dot{f}^{2}%
}{f}+H\dot{f}\right) -\left( N\ddot{f}-\ddot{N}f\right) \right] -\frac{1}{%
16\pi ^{2}qf^{2}}\left( N\dot{f}+Hf\right) .  \label{in4}
\end{equation}%
In the conventional inflationary scenario, the $\epsilon _{H}$ grows as well
$d\epsilon _{H}/dN=2\epsilon _{H}\left( \eta -\epsilon _{H}\right) $. In
this case, the spectral index of perturbations is equal to $%
n_{s}=1-6\epsilon _{H}+2\eta $ \cite{L15}. We can estimate the spectral
index by the ratio%
\begin{equation}
n_{s}=1+\frac{2H_{f}}{H}+\frac{H+NH_{f}}{4\pi ^{2}qf}\left[ 1-\frac{4\pi
^{2}q\left( N\ddot{f}-\ddot{N}f\right) }{H\left( H+NH_{f}\right) ^{2}}\right]
,  \label{in5}
\end{equation}%
One can define a Hubble parameter $H_{f}=\dot{f}/f$\ associated with $f$.
The Planck CMB data implies that $n_{s}\approx 0.965\pm 0.004$ \cite{L2}. We
introduce the effective expansion rate $H_{eff}$ \cite{R24,R25}, which is
evaluated as follows:%
\begin{equation}
H_{eff}=H+NH_{f}.  \label{in6}
\end{equation}%
The temperature remains nearly unchanged at the time of formation of the
PBHs \cite{L17,L18}, i.e. from Eq. (\ref{p4}) we have $\ddot{N}=\left( \frac{%
2\pi \hbar }{k_{B}T}\right) \ddot{f}$. Thus we obtain $N\ddot{f}=\ddot{N}f$.
Recall that the spectral index is given by Eq. (\ref{in5}) and we use the $N%
\ddot{f}=\ddot{N}f$, we find the relationship between the spectral index and
the current expansion rate:\textbf{\ }%
\begin{equation}
n_{s}=1-\frac{2}{N}+\left( \frac{2}{H_{0}N}+\frac{1}{4\pi ^{2}qf}\right)
H_{eff,0}.  \label{NS}
\end{equation}%
From Eqs. (\ref{p5}) and (\ref{NS})\textbf{,} we find $n_{s}\approx 1+\frac{%
2H_{eff}}{H_{0}N}-\frac{2}{N}+\frac{4GM}{c^{3}}\frac{H_{eff}}{qN}.$ We note
the presence in this expression the presence of the Schwarzschild radius. We
notice that $H_{eff}$ includes the number of e-folds. In other words, the
term $H_{f}$ affects the evolution of $H_{eff}$. This behavior is very
similar to the one discussed for the Hubble tension \cite{R18,R19}. It is
useful to calculate the value of $H_{f,0}$ from the data we have available, $%
AsH_{eff,0}\approx \left( 73.24\pm 1.74\right) km.s^{-1}.Mpc^{-1}$ based on
type SNeIa observations from the HST \cite{L1} and $H_{0}\approx \left(
67.27\pm 0.66\right) km.s^{-1}.Mpc^{-1}$ by CMB measurements, for $N=60$, we
can calculate $H_{f,0}$ to be $H_{f,0}\approx \left( 0.0995\pm 2.4\right)
km.s^{-1}.Mpc^{-1}$ by Eq. (\ref{in6}). Next we want to elaborate on the
validity of Eq. (\ref{in6}) by numerical calculation which is given in Table %
\ref{t3}. Recall that the spectral index is given by Eq. (\ref{in5}) for $N%
\ddot{f}\approx \ddot{N}f.$\textbf{\ }According to the numerical results and
from Eq. (\ref{in6}), we have $H_{0}N\ll 8\pi ^{2}qf$. So,
\begin{equation}
n_{s}=1+\frac{2}{N}\left( \frac{H_{eff,0}}{H_{0}}-1\right) .  \label{inn7}
\end{equation}%
In the limit $H_{eff,0}\approx H_{0}$, one gets $n_{s}\approx 1$.
\begin{table}[H]
\begin{center}
\begin{equation*}
\begin{tabular}{cccccc}
\hline
Measurements & $H_{eff,0}(km.s^{-1}.Mpc^{-1})$ & $H_{0}(km.s^{-1}.Mpc^{-1})$
& $N$ & $n_{s}$ & $n_{s}^{moy}$ \\ \hline\hline
CMB with Planck \cite{L2} & $67.27\pm 0.60$ & $67.27$ & $60$ & $%
1_{-0.0002}^{+0.0002}$ & $1$ \\
CMB without Planck \cite{AL1} & $68.8\pm 1.5$ & $67.27$ & $60$ & $1.0015$ & $%
1.0015$ \\
No CMB with BBN \cite{AL2} & $68.5\pm 2.2$ & $67.27$ & $60$ & $%
1_{-0.0004}^{+0.0016}$ & $1.0006$ \\
CMB lensing \cite{AL3} & $70.6_{-5.0}^{+3.7}$ & $67.27$ & $60$ & $%
1_{-0.0008}^{+0.0034}$ & $1.001$ \\
Cepheids - SNIa \cite{AL4} & $73.2\pm 1.3$ & $67.27$ & $60$ & $%
1_{-0.0022}^{+0.0035}$ & $1.002$ \\
TRGB -SNIa \cite{AL5} & $72.1\pm 2.0$ & $67.27$ & $60$ & $%
1_{-0.0014}^{+0.0033}$ & $1.002$ \\
Miras -SNIa \cite{AL6} & $73.2\pm 4.0$ & $67.27$ & $60$ & $%
1_{-0.0009}^{+0.0049}$ & $1.002$ \\
Masers \cite{AL7} & $73.9\pm 3.0$ & $67.27$ & $60$ & $1_{-0.0017}^{+0.0047}$
& $1.003$ \\
HII galaxies \cite{AL8} & $71.0\pm 3.5$ & $67.27$ & $60$ & $%
1_{-0.0001}^{+0.0035}$ & $1.001$ \\
GW related \cite{AL9} & $73.4_{-10.7}^{+6.9}$ & $67.27$ & $60$ & $%
1_{-0.0022}^{+0.0064}$ & $1.003$ \\ \hline
\end{tabular}%
\end{equation*}%
\end{center}
\caption{Numerical estimate of the values of $n_{s}$ from Eq. ({\protect\ref%
{inn7}}) with Hubble constant $H_{eff,0}$ through direct and indirect
measurements \protect\cite{L16}. We take $H_{0}=67.27km.s^{-1}.Mpc^{-1}$ as
the reference. Note that the values of $n_{s}^{moy}$\ represent the average
value of $n_{s}$.}
\label{t3}
\end{table}
In the radiation-dominated era ($q=\frac{1}{2}$, $N=60$) at the frequency $%
f=5.5nHz$, the current effective rate is
\begin{equation}
H_{eff,0}=30\left( n_{s}-0.96\right) H_{0}.  \label{in7}
\end{equation}%
For $H_{0}\approx 67.27km.s^{-1}.Mpc^{-1}$ and $n_{s}^{\ast }\approx 0.996$,
we find $H_{eff,0}\approx 73.24km.s^{-1}.Mpc^{-1}$. This spectral index
approves the constraint $n_{s}>0.9$ \cite{R21}. While Blais et al. \cite{R20}
find that the mass fluctuation is reduced by $34\%$ for scale-free power law
primordial spectra, and the spectral index is in the range $1\leq n_{s}\leq
1.3$. For that we widen the domain of existence by $n_{s}$: $0.996\leq
n_{s}\leq 1.3$. The critical value of the spectral index is $n_{s}^{0}=0.96$%
\ which is compatible with the Planck 2018 results ($n_{s}^{0}\approx
0.965\pm 0.004$) \cite{L2}, also with $n_{s}^{0}\approx 0.9569$ found in the
inflationary models of PBH with an inflection point \cite{R22,R23}.

\section{PBHs formation and evaporation}

\begin{figure}[H]
\centering\includegraphics[width=11cm]{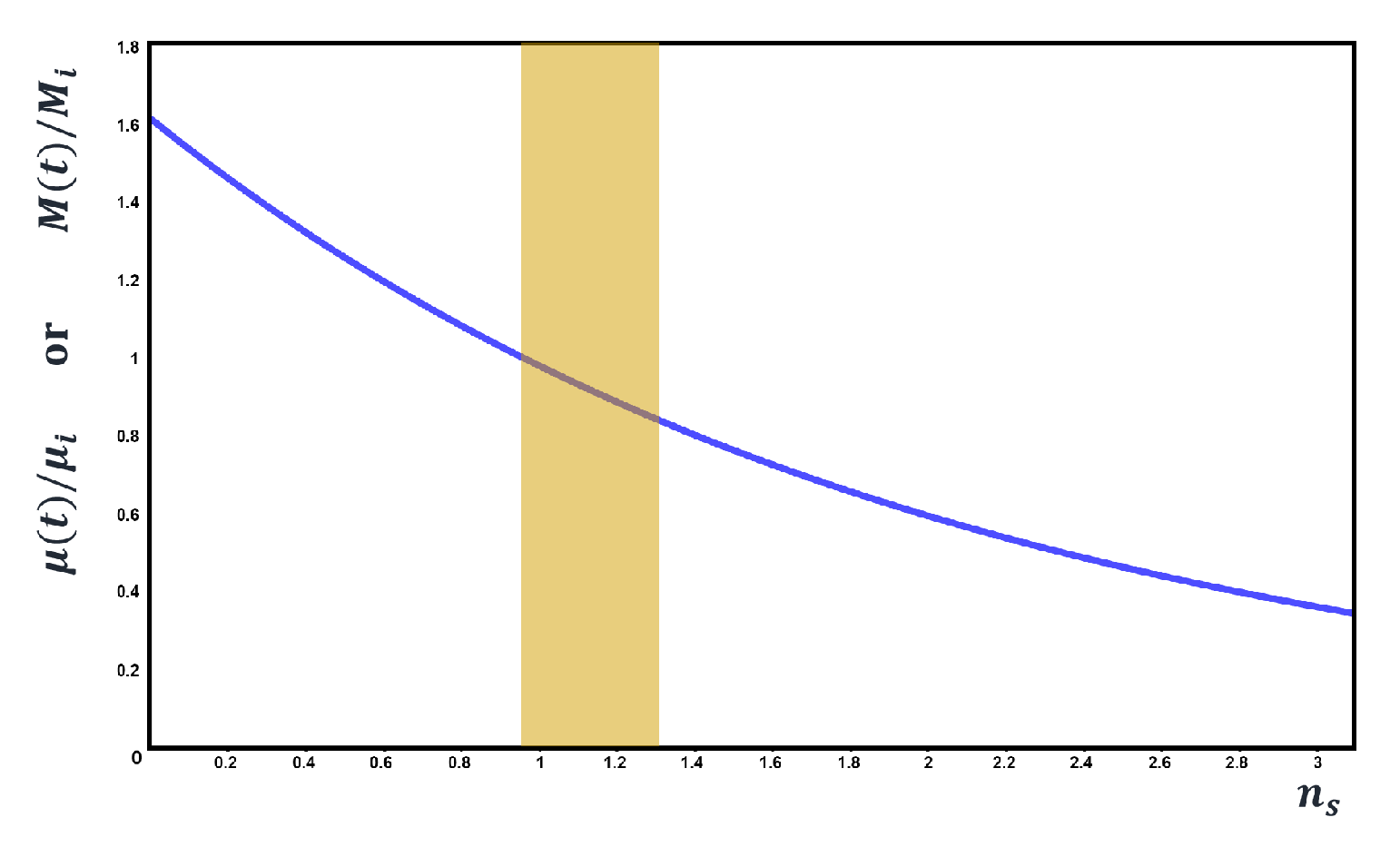}
\caption{Mass ratio $M(n_{s})/M_{i}$ (or the chemical potential ratio $%
\protect\mu (n_{s})/\protect\mu _{i}$ Eq. (\protect\ref{e4})) as a function
of the number of the spectral index $n_{s}$. The orange area represents the
validity interval of $n_{s}$.}
\label{f1}
\end{figure}

In this section, we are interested in studying the formation of PBHs and
their subsequent evaporation. The initial PBH mass $M$ is close to the
Hubble horizon mass $M_{H}$ and therefore given by $M=\gamma M_{H}$ \cite%
{PBH02} where $\gamma $ is a factor that depends on the gravitational
collapse nature of the black hole and the radiation quantity in the early
Univers. The Hubble horizon mass is%
\begin{equation}
M_{H}=\frac{4\pi R_{H}^{3}}{3}\rho ,  \label{q1}
\end{equation}%
where $R_{H}$ is the black hole horizon radius and $V=\frac{4\pi R_{H}^{3}}{3%
}$ is the thermodynamic volume. The mass fraction of PBHs at the time of
formation is%
\begin{equation}
\beta =\frac{\rho _{PBH,i}}{\rho _{r,i}},
\end{equation}%
where $\rho _{PBH,i}$ is the PBHs\ energy density and $\rho _{r,i}$\ is the
background radiation density at the PBH formation epoch (index $i$). The
current density parameter $\Omega _{PBH}$ of the PBHs is related to the
initial collapse fraction $\beta $ by%
\begin{equation}
\Omega _{PBH}=\beta \left( 1+z\right) \Omega _{CMB},  \label{q2}
\end{equation}%
where $\Omega _{CMB}\approx 5\times 10^{-5}$ is the density parameter of the
CMB and $z$ is the redshift. The $(1+z)$ factor can be understood as arising
for the radiation density scales as $(1+z)^{4}$, whereas the PBH density
scales as $(1+z)^{3}$. The PBHs have a mass of order $M\sim \frac{tc^{3}}{G}$%
. We introduce the fraction $f_{PBH}(M)$ of the halo in PBHs for $\Omega
_{CDM}\approx 0.264$ \cite{L2} by $f_{PBH}\left( M\right) =\frac{\Omega
_{PBH}}{\Omega _{CDM}}\approx 3.79\Omega _{PBH}$. A constraint on $\Omega
_{PBH}$ for $M>10^{15}g$ is $\Omega _{PBH}<\Omega _{CDM}$, this condition
describes non-evaporating PBHs. For the $\gamma $-ray background limit, we
have $\beta \left( 10^{15}g\right) \lesssim 10^{-29}$. The PBHs evaporate at
the epoch of cosmological nucleosynthesis if they are characterized by mass $%
M\approx 10^{10}g$, temperature $T_{PBH}\approx 1TeV$ and lifetime $\tau
\approx 10^{3}s$. These PBHs have an effect of these evaporations on the
BBN. Using the relation $\rho =3sT/4$ \cite{PBH02}, the fraction of the
Universe's mass in PBHs at their formation time is then related to their
number density $n_{PBH}$:%
\begin{equation}
\beta =\frac{4M}{3T}\frac{n_{PBH}}{s}=\frac{4n_{PBH}}{3k_{B}^{2}T_{0}}M.
\label{qq5}
\end{equation}%
The black hole mass $M$\ is identified with the thermodynamic enthalpy $%
\mathcal{H}$ with $d\mathcal{H}=dM$. The black hole first law reads%
\begin{equation}
dM=Tds+VdP.  \label{qq1}
\end{equation}%
From Eq. (\ref{p3}) and Eq. (\ref{qq1}) we obtain%
\begin{equation}
\rho =\frac{dM}{dV}=k_{B}T\frac{dN}{dV}+V\frac{dP}{dV}.  \label{qq2}
\end{equation}%
Using $VdP=-PdV$. As a consequence,%
\begin{equation}
\rho c^{2}+P=nk_{B}T,  \label{qq3}
\end{equation}%
where $n=dN/dV$ is the density of the e-folds number. Using this relation
the e-folds number density is $n=\frac{\rho c^{2}+P}{k_{B}T}$. Certain phase
transitions lead to a sudden reduction in pressure, which enhances the
formation of PBH\textrm{\ }\cite{R11}. In a stable fluid sphere the equation
of state parameters $\omega \equiv P/\rho c^{2}$ should be positive. The
pressure $P$ and density $\rho $ should be positive inside the stars and
should satisfy the energy conditions $\rho c^{2}+P\geq 0$ \cite{R4}. For
this purpose, the number density $n$ should be positive. As mentioned
before, the PBH mass is given by the horizon mass when the fluctuations
reenter the horizon. With this, the energy $Mc^{2}=\gamma V\rho c^{2}$ Eq. (%
\ref{q1}) is connected with the Smarr formula, which now reads%
\begin{equation}
\gamma ^{-1}Mc^{2}+PV=T\frac{ds}{d\ln V},
\end{equation}%
and for $n=N/V$, we can recover the standard form of Smarr formula%
\begin{equation}
\gamma ^{-1}Mc^{2}+PV=Ts.
\end{equation}%
We notice the difference between the black holes which verified the Smarr
formula \cite{R10}. Using the general Smarr formula in 4-dimension brought
in \cite{R10}, we estimate that $\gamma \sim 2$. The mass loss rate of a PBH
is obtained from Eq. (\ref{p5}) as
\begin{equation}
\frac{dM}{dt}=-\frac{c^{3}}{16\pi ^{2}Gf}\left( H+NH_{f}\right) <0.
\label{qq6}
\end{equation}%
This mass loss is due to Hawking evaporation. Using Eqs. (\ref{p5},\ref{qq6}%
), we obtain
\begin{equation}
\frac{d\ln M}{dt}=-\left( \frac{H}{N}+H_{f}\right) .  \label{qq7}
\end{equation}%
The time evolution of the PBH mass with initial mass $M_{i}$ formed at $%
t_{i} $ is evaluated by relation%
\begin{equation}
M(t)=M_{i}e^{-\left( \frac{H}{N}+H_{f}\right) t}.  \label{qq8}
\end{equation}%
According to this expression, we see the total disappearance of PBHs for $%
t\rightarrow \infty $. Note that the PBHs would have completely evaporated
until today if their mass is lighter than $M=10^{15}g$. At the end of
inflation ($N=0$), the PBH mass vanishes. Some PBHs generated before
inflation are diluted to negligible density \cite{PBH02}. This agrees with
Eqs. (\ref{qq8}), when $N=60$, we have $M(t)\neq 0$. The effective Hubble
rate can be obtained from Eq. (\ref{qq8}):%
\begin{equation}
H_{eff}=-N\frac{d\ln M}{dt}.  \label{qq9}
\end{equation}%
In order to estimate precisely the fluctuations generated in the Hubble
rate, we note that the PBHs evaporation influences the evolution of $H_{eff}$%
. Recall that the spectral index is given by Eq. (\ref{in5}) and by Eq. (\ref%
{in7}) we calculate mass ratio as:%
\begin{equation}
M(n_{s},t=t_{0})/M_{i}=e^{\frac{0.96-n_{s}}{2}}.  \label{qq10}
\end{equation}%
where $t_{0}=1/H_{0}\approx 4.22\times 10^{17}s$, represents the age of the
Universe. For the initial PBH mass $M_{i}=10^{15}g$ \cite{L11} and $%
n_{s}^{\ast }\approx 0.996$ we find $M^{\ast }=9.8216\times 10^{14}g$. In
Fig. (\ref{f1}), we plot the ratio $M/M_{i}$ as a function of $n_{s}$.
For a constant $N$, we introduce the effective scale factor:%
\begin{equation}
a_{eff}=a_{eff,0}\left( \frac{M(t)}{M_{i}}\right) ^{-N}.  \label{qq11}
\end{equation}%
The factor $a_{eff,0}$\ represents the value of $a_{eff}$\ at the end of
inflation ($N=0$). Surprisingly, the $a_{eff}$ is related to the PBH mass,
which can generate a big difference in the measurements of the Hubble rate
in the primordial universe and by this fact, we propose that PBHs which are
formed from primordial perturbations, generate the Hubble tension.

\section{Chemical potential and central charge}

\begin{figure}[H]
\centering\includegraphics[width=11cm]{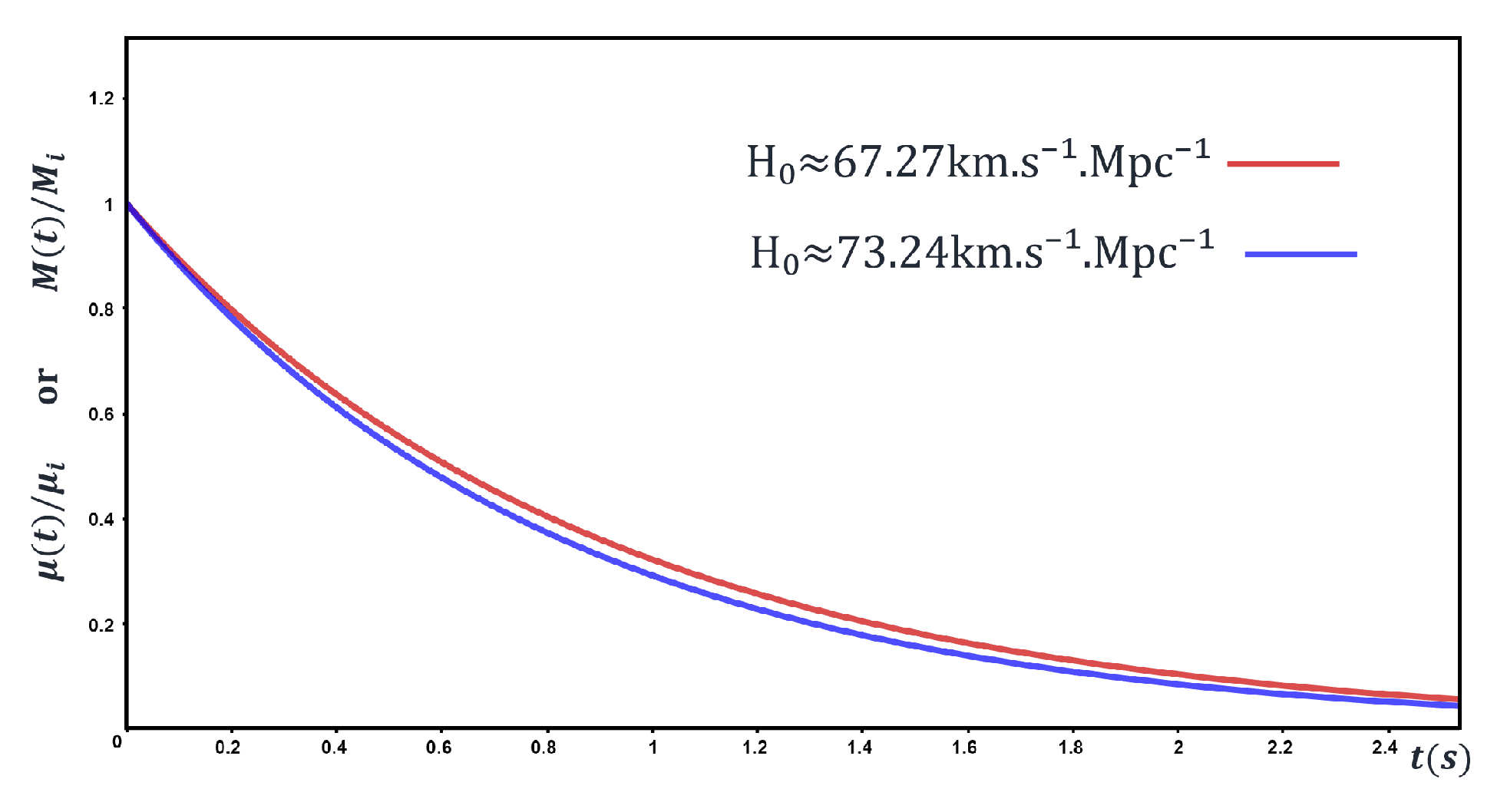}
\caption{Curves of $M/M_{i}$ or $\protect\mu /\protect\mu _{i}$, with
respect to time. For $N=60$.}
\label{f2}
\end{figure}

Recently, in the context of the gauge/gravity duality, M. R. Visser \cite{R3}
proposed a novel holographic central charge $C=L^{d-2}/l_{p}^{d-2}$ in which
$l_{p}$ is the Planck length, which has attracted great attention. He finds
that the Euler equation is dual to the generalized Smarr formula for black
holes in the presence of a negative cosmological constant.\textbf{\ }Thus,
we replace the entropy value that depends on $N$\ Eq. (\ref{p3}) in Eq.(\ref%
{qq1}) which gives
\begin{equation}
dM=\frac{k_{B}T}{2N}dN^{2}+VdP.  \label{e1}
\end{equation}%
We note that N is constant at the time of formation of PBHs which is equal
to 60, while during the evolution of PBHs the value of N changes according
to this relation. This expression gives an interpretation of the number of
e-folds as the number of microscopic degrees of freedom for the PBH. This
equation involves an AdS space-time and a dual CFT \cite{R1,R2} of the basic
structure of Visser's framework \cite{R3}. We have $C\equiv N^{2}$, which
corresponds to the result found by the conformal symmetry in $SU(N)$ \cite%
{R3}. Thus we can uniformly rewrite Eq. (\ref{e1}) as
\begin{equation}
dM=\mu dC+VdP.  \label{e2}
\end{equation}%
The volume $V$\ and the chemical potential $\mu $\ are the conjugate
quantities of the pressure $P$\ and the central charge $C$, respectively.
This equation mixes the volume and boundary description of a black hole
because $C$\ is the central boundary charge and $P$\ is the volume pressure.
In d-dimensional CFT there are several candidates for the central charge of
e-folds $C$. In the dual CFT, the chemical potential $\mu $ is the conjugate
quantity of the central charge, this corresponds to including $\left( \mu
,C\right) $ as a new pair of conjugate thermodynamic variables \cite{R3}.
This leads to the following relation%
\begin{equation}
\mu \left( T\right) =\frac{k_{B}T}{2N}.  \label{e3}
\end{equation}%
The solution of black hole chemical potential found in \cite{R1} reduces to
this solution of PBHs. We notice that the chemical potential in Eq. (\ref{e3}%
) is linked to the thermal energy of e-folds (equipartition theorem: $%
E_{th}=Nk_{B}T/2$). This leads to $E_{th}=\mu C=\pi \hbar f$, i.e., thermal
energy exchange occurs between PBHs during inflation with the emission of
gravitational waves of frequency $f$. Making use of Eq. (\ref{qq8}), another
suggestive expression for the chemical potential could be%
\begin{equation}
\mu \left( t\right) =\mu _{i}e^{-\left( \frac{H}{N}+H_{f}\right) t},
\label{e4}
\end{equation}%
where $\mu _{i}=\frac{M_{i}q}{N^{2}}$. Similarly, the chemical potential
becomes zero as inflation comes to an end through the condition $N=0$. From
Eq. (\ref{qq10}) we get $\mu (n_{s})/\mu _{i}=e^{\frac{0.96-n_{s}}{2}}$ (see
Fig. (\ref{f1})). The variation of the chemical potential with time is
plotted in Fig. (\ref{f2}). Using $n_{s}^{\ast }\approx 0.996$, we find $\mu
/\mu _{i}\approx 0.98216$ which is in good agreement with the result of $\mu
/\mu _{i}$\ for CMB with Planck \cite{L2} in Table \ref{t4}.
\begin{table}[H]
\begin{center}
\begin{equation*}
\begin{tabular}{cccc}
\hline
Measurements & $H_{eff,0}(km.s^{-1}.Mpc^{-1})$ & $n_{s}$ & $\mu /\mu _{i}$
\\ \hline\hline
CMB with Planck \cite{L2} & $67.27\pm 0.60$ & $1$ & $0.9801$ \\
CMB without Planck \cite{AL1} & $68.8\pm 1.5$ & $1.0015$ & $0.9794$ \\
No CMB with BBN \cite{AL2} & $68.5\pm 2.2$ & $1_{-0.0004}^{+0.0016}\left(
+0.0006\right) $ & $0.9799$ \\
CMB lensing \cite{AL3} & $70.6_{-5.0}^{+3.7}$ & $1_{-0.0008}^{+0.0034}\left(
+0.001\right) $ & $0.9797$ \\
Cepheids - SNIa \cite{AL4} & $73.2\pm 1.3$ & $1_{-0.0022}^{+0.0035}\left(
+0.002\right) $ & $0.9792$ \\
TRGB -SNIa \cite{AL5} & $72.1\pm 2.0$ & $1_{-0.0014}^{+0.0033}\left(
+0.002\right) $ & $0.9792$ \\
Miras -SNIa \cite{AL6} & $73.2\pm 4.0$ & $1_{-0.0009}^{+0.0049}\left(
+0.002\right) $ & $0.9792$ \\
Masers \cite{AL7} & $73.9\pm 3.0$ & $1_{-0.0017}^{+0.0047}\left(
+0.003\right) $ & $0.9787$ \\
HII galaxies \cite{AL8} & $71.0\pm 3.5$ & $1_{-0.0001}^{+0.0035}\left(
+0.001\right) $ & $0.9797$ \\
GW related \cite{AL9} & $73.4_{-10.7}^{+6.9}$ & $1_{-0.0022}^{+0.0064}\left(
+0.003\right) $ & $0.9787$ \\ \hline
\end{tabular}%
\end{equation*}%
\end{center}
\caption{Numerical estimate of the minimal variation of $\protect\mu /%
\protect\mu _{i}$\ according to the observations.}
\label{t4}
\end{table}
In the radiation-dominated era ($q=\frac{1}{2}$) we have $\mu _{i}=\frac{%
M_{i}}{2N^{2}}$. While, in the matter-dominated era ($q=\frac{3}{2}$) we
have $\mu _{i}=\frac{2M_{i}}{3N^{2}}$. Variation of the Gibbs free energy of
PBHs that is held at constant pressure and temperature is $dG=\mu dN$. The
Gibbs free energy is at its minimum if $dG=\mu dN=0$, as a consequence, the
number $N$ is constant (for example $N=60$). With this, we discuss the
chemical potential conjugate to the number of e-folds. Alternatively, the
Hawking-Page temperature is obtained via the vanishing of the Gibbs free
energy.

\section{Conclusions}

This manuscript offers an explanation for the observed gravitational-wave
background signal detected by pulsar timing arrays. The proposed model
suggests that the stochastic gravitational waves originate from primordial
black holes. We extensively explore the implications of NANOGrav data within
the context of inflationary scenarios and draw comparisons between this
model and primordial perturbations. In our study, we present a frequency
relation associated with the stochastic gravitational wave background.
Additionally, we calculate the mass, temperature, and entropy corresponding
to the GW frequency $f$. By establishing a connection between the horizon
mass and the comoving frequency of the perturbation, we derive a
relationship between the PBH mass $M$ and the GW frequency $f$ as $M\propto
1/f$. Through analytical and numerical analyses, we demonstrate that the
frequency $f$ aligns well with the constraints on the mass range of $%
10^{-5}kg-10^{50}kg$. Furthermore, we derive a probability distribution
relation as a function of frequency. The obtained relation aligns well with
the estimated sensitive volume for GW source populations at redshift,
indicating a strong agreement. Moreover, we have established a correlation
between entropy and the number of e-folds. Through the expression of
temperature in terms of $N$ (number of e-folds) and frequency, we have
discovered that primordial black holes (PBHs) exhibit extremely low
temperatures, a characteristic that supports the cold dark matter model.
Additionally, we have introduced a new parameter derived from this
frequency. The presence of this parameter signifies the rate of mass loss
for an evaporating black hole. It is possible that this parameter could
offer a resolution to the problem of the Hubble tension. From this
perspective, the effective Hubble constant $H_{eff,0}$ can be expressed in
relation to the number of e-folds $N$ and the Hubble constant $H_{0}\approx
67.27km.s^{-1}.Mpc^{-1}$, and subsequently in terms of the spectral index $%
n_{s}$. Numerical calculation led to values: $n_{s}\approx 0.996$ and $%
H_{eff,0}\approx 73.24km.s^{-1}.Mpc^{-1}$, which is in good agreement with
the observational data. Subsequently, we suggested the domain $n_{s}$: $%
0.996\leq n_{s}\leq 1.3$ for the spectral index. We suggested new
interpretations for the thermodynamic parameters such as chemical potential
and the central charge. We interpreted the number of e-folds as the number
of microscopic degrees of freedom for the PBH and have shown that the
temperature undergoes Wien's displacement law. Development of the proposed
approach gives ground for a principally new scenario of the formation and
evaporation of the primordial black holes in the early Universe. This opens
a new window to study the relationship between Hubble tension and stochastic
gravitational waves.

\end{document}